\documentclass[preprint,authoryear,12pt]{chrArticle}

\usepackage{graphicx}
\usepackage{units}
\usepackage{amssymb}
\usepackage{setspace}
\usepackage[margin=1in]{geometry}
\usepackage{float}

\usepackage{amsmath, amstext, amsfonts, mathrsfs, amssymb}
\usepackage{rotating,xcolor}

\usepackage{christiansTexDefs}
                                                                                                                             
\journal{International Journal of Multiphase Flow}

\begin{document}

\begin{frontmatter}

\title{A priori and a posteriori analysis of models for Large-Eddy simulation of particle-laden flow}

\author{ Christian Gobert and Michael Manhart}
\ead{christian.gobert@mytum.de}
\address{TU M\"unchen, Fachgebiet Hydromechanik,
Arcisstr. 21, D-80333 M\"unchen, Germany}

\begin{abstract}
In Large-Eddy simulation of particle-laden flow, the effect of the unresolved scales
on the particles needs to be modelled. In this work we analyse three very promising models, namely
the approximate deconvolution method (ADM) which was proposed for particle-laden flow independently by Kuerten (Phys. Fluids 18, 2006) and 
Shotorban and Mashayek (Phys. Fluids 17, 2005)
and two stochastic models, proposed by Shotorban and Mashayek (J. Turbul. 7, 2006) and Simonin et al. (Appl. Sci. Res. 51, 1993). We present results from
a priori and a posteriori analysis of these
models in isotropic turbulence at $Re_\lambda=52$. This data allows for a direct quantitative comparison of the models. The analysis shows that ADM
always leads to improved statistics but that even for high Stokes numbers, the rate of dispersion is not predicted correctly by ADM.
Concerning the stochastic models, we found that
with the correct choice of model parameters,
the models perform well at small Stokes numbers. On the other hand, at high Stokes numbers
the stochastic models show significant errors such that
it may be recommendable to neglect the small scale effects instead of using one of the stochastic models.

\end{abstract}


\end{frontmatter}

\section{Introduction}
\label{chapAssessment}

Large-Eddy Simulation (LES) has become an important tool for the simulation of turbulent flow. State of the art methods provide reliable
results and are capable to tackle application relevant challenges. One crucial component for LES is the correct choice of a turbulence model, i.e.,
a model for the effect of the unresolved subgrid scales (SGS) on the resolved scales. Such models are herein referred to as {\it fluid-LES models}.

For Large-Eddy Simulation of particle laden flow, an additional model for the
effect of the unresolved scales on the particles is needed, referred to as {\it particle-LES models}. 
The works of \cite{Yamamoto01,Ar99,Kuerten05,citeulike:1477237,citeulike:3487195,citeulike:3818012}
show that neglection of small scale effects is not an option.

Most particle-LES models were developed in a Eulerian-Lagrangian framework, i.e., the carrier fluid flow is computed by solving the Navier--Stokes equations
and the particles are computed by tracing single particles through the domain. Then, modelling reduces to reconstruction of small scale effect
on a single particle.

On this basis, a large number of models was proposed. Among these are
for example the models of \cite{Simonin93,Wang96,citeulike:3191448,Kuerten06,citeulike:839184,e_Gobert06a,citeulike:3818015,
citeulike:3817997,citeulike:3836482,citeulike:6598002,citeulike:6627481}, just to mention a few.
Most of these models are stochastic models, often
obtained by extending models which were originally developed for inertia free particles
in the context of Reynolds Averaged
Navier--Stokes (RANS) simulations, such as the generalised Langevin model by \cite{citeulike:3817826}.

A deterministic alternative is the approximate deconvolution method (ADM)
 for particle laden flows \cite[see][]{Kuerten06,citeulike:3818015,citeulike:1476706}. 
 ADM is based on an approximate inversion of the LES filter and was originally developed in
 a Eulerian context.

The present study focusses on three very promising particle-LES models, namely the stochastic
models proposed by \cite{citeulike:3191448} and \cite{Simonin93} and ADM as proposed by \cite{Kuerten06} and \cite{citeulike:1476706}.
For all three models, the respective authors present some results on the accuracy of their models. Their findings are summarised as follows.

\cite{Kuerten06} analysed ADM in particle-laden turbulent channel flow at a Reynolds number based on friction velocity of
$Re_\tau=150$. He conducted an a posteriori analysis for particles with Stokes numbers of $St=1$, 5 and 25, based on the viscous time scale.
His results show that ADM significantly improves rms values of the wall normal component of the particle velocity. The improvement is 
greater for high Stokes number than for low Stokes number. 
In addition,
\cite{citeulike:1476706} found that in a turbulent shear layer,
ADM improves particle dispersion.

\cite{citeulike:1476706} analysed their Langevin-based model in decaying
isotropic turbulence and found that for small Stokes numbers ($St \le 2.5$ based on the Kolmogorov time scale at initialisation) 
the model leads to correct particle dispersion 
whereas at higher Stokes number significant deviations can be observed. 

\cite{citeulike:1666260} analysed the model of \cite{Simonin93} in forced isotropic
turbulence and found that the model leads to correct kinetic energy for the particles. However, their simulations are restricted to
$St\le 5$, based on the Kolmogorov time scale. The present study shows that at higher Stokes numbers the model does not perform very well.

Concluding, all published results were obtained on different configurations and are therefore not comparable.
In particular, for the Langevin-based models 
only data at small Stokes numbers is published. 
For all models, the available data density over the Stokes number range is not satisfactory. Data rather correspond
to probes at specific Stokes numbers but from this data no Stokes number dependent behaviour of the models can be deduced. 

The present study aims at a 
clarification of that issue by
providing data which allows a direct comparison of these three particle-LES models
on a broad range of Stokes numbers. The data density on the Stokes number range is sufficiently high to allow the deduction of a Stokes number dependence.
The testcase is isotropic turbulence at $Re_\lambda=52$. All three models were originally developed such that
they should perform well in that testcase but we will show that even by tuning the model constants, the models do not
always perform well. Actually in some cases better results are obtained by neglecting SGS effects than using one of the stochastic models.

This paper is organised as follows. Sections \ref{secNumCarr} and \ref{secDPS} contain a description of the
numerical methods used to compute flow and particle dynamics. Statistics of the single
phase simulations are also presented in section \ref{secNumCarr}. 
In section \ref{secModels}, the three particle-LES models under consideration are presented and 
section \ref{secNumAssess} contains results of an a priori
and an a posteriori analysis of the models.

\section{Numerical Simulation of the carrier flow}
\label{secNumCarr}

In the present work we analyse 
particle dynamics in forced isotropic turbulence by DNS and LES. 
For the simulation of the carrier fluid, we use a second order Finite-Volume method together with a
third order Runge-Kutta scheme proposed by \cite{Williamson80} for advancement in time. The
conservation of mass is satisfied by solving the Poisson
equation for the pressure using an iterative
solver proposed by \cite{citeulike:3040144}.
 More details on the flow solver can be found in \cite{e_Manhart04}. 

The flow is driven using a slightly modified version of the
deterministic forcing scheme proposed by \cite{Sullivan94}.
Sullivan et al. propose a forcing scheme where the energy in the spectral modes below a
certain wave number $\kappa_1$ is held constant. We additionally imposed a lower bound for the forced
wavenumbers, i.e., 
only the modes in a given range $[\kappa_0, \kappa_1]$ are forced. 
The Reynolds number in our simulations is always
$Re_\lambda= 52$,
based on the transverse
Taylor microscale $\lambda$ and the rms value of one (arbitrary) component of the
fluctuations $u_{\mt{rms}}$.

In all computations the flow was solved in a cube on a staggered Cartesian equidistant grid. The size of the computational
box and the cell width was chosen such that all scales are resolved, based on the criteria stated by \cite{citeulike:1666191},
cf. table \ref{tabDNSData}.

\begin{table}[h]
\begin{center}
\caption{\label{tabDNSData}Simulation parameters and Eulerian statistics from DNS of forced isotropic turbulence.
}
\begin{tabular}{ll}
\multicolumn{2}{c}{DNS} \\
\hline
\small
$Re_\lambda$                       & 52               \\
\hline
Number of grid points $N$                                & $256^3$          \\
range of forced wavenumbers $[\kappa_0, \kappa_1] $     & $[0.514, 1.54] / \lambda$ \\
integral length scale $L_f$                    & $2.00 \lambda$             \\
time scale of energy containing eddies $k_f / \epsilon$    & $5.15 \lambda/u_{rms}$             \\
Kolmogorov length scale $\eta_K$                 & $0.070  \lambda$            \\
Kolmogorov time scale $\tau_K$           & $0.248 \lambda/u_{rms}$            \\
length of computational box $L$                        & $11.9 L_f$            \\
cell width $\Delta x$                 & $1.34 \eta_K$            \\
filter width $\Delta$                  & $7 \Delta x$                \\
kinetic energy of the filtered field $\hat{k}_f$                  & $0.87  / k_f$             \\
\end{tabular}
\end{center}
\end{table}

The particle-LES models were assessed by a priori and a posteriori
analysis. For the a priori analysis, we filtered the DNS field $\bf u$ by a box filter $\cal G$ 
with filter width $\Delta =7 \Delta x$, $\Delta x$ being the DNS cell width.
The filtered DNS was sampled on a correspondingly coarse grid, resulting in a field ${\cal G}{\bf u}$ which is comparable to an LES field.
The kinetic energy of the filtered field $\hat{k}_f=\left< {\cal G}{u}_i^2 \right> / 2$ is 87\% of the
energy of the unfiltered field $k_f=\left< u_i^2 \right> / 2$, cf. table \ref{tabDNSData}. $\left< \cdot \right>$ denotes spatial and
temporal averaging.

In the a priori analysis,
${\cal G}{\bf u}$ was used as input for the particle-LES models. Then, the models 
were assessed with respect to the difference in statistics obtained from unfiltered DNS and filtered DNS with particle-LES model.

\begin{figure}[H]
\includegraphics[bb=87 5 706 320, width=8.2cm]{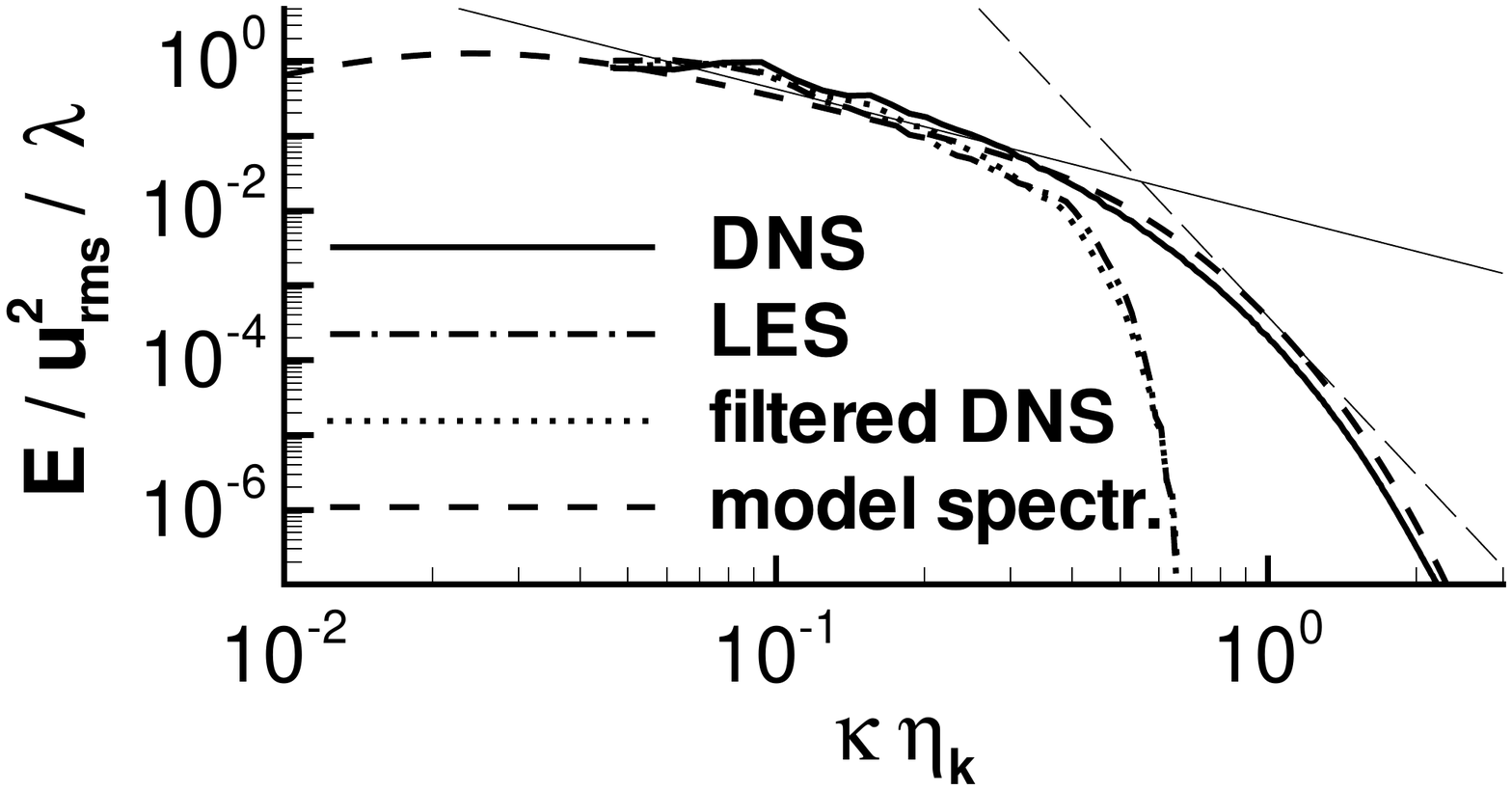}
\caption{\label{figSpectr}Instantaneous energy spectrum functions together with lines proportional to $\kappa^{-5/3}$ and $\kappa^{-7}$.}
\end{figure}

For the a posteriori analysis, ${\cal G}{\bf u}$ was computed by LES. 
As fluid-LES model we used
the Lagrangian dynamic Smagorinsky model proposed by 
\cite{Meneveau96}. 
The forcing parameters for LES
were chosen in the same way as for DNS, i.e., the energy contained in the range $[\kappa_0,\kappa_1]$ is equal in LES and DNS.
Beyond $\kappa_1$, the energy in LES is lower than in DNS due to the different grids and the fluid-LES model.
With our choice of the grid,
the kinetic energy resolved by LES $\bar{k}_f$ 
is approximately equal to the kinetic energy of the
filtered DNS field $\hat{k}_f$, cf. table
\ref{tabLESData}. Instantaneous energy spectra $E(\kappa)$ from DNS and LES are plotted in figure \ref{figSpectr}. 
In addition, a model spectrum proposed by \cite{citeulike:1666191} is shown.
All data were made dimension free by normalising with DNS quantities.

\begin{table}[h]
\begin{center}
\caption{\label{tabLESData}Parameters for LES of forced isotropic turbulence.}
\begin{tabular}{ll}
\multicolumn{2}{c}{LES} \\
\hline
$Re_\lambda$ &  52 \\
\hline
Number of grid points $N$  & $42^3$ \\
cell width $\Delta x$ & $0.567 \lambda$  \\
time scale of energy containing eddies $\bar{k}_f / \bar{\epsilon}$    & $10.40 \lambda/u_{rms}$             \\
resolved kinetic energy $\bar{k}_f$ & $0.86 k_f$ \\
\end{tabular}
\end{center}
\end{table}

\section{Discrete particle simulation}
\label{secDPS}

In this study we consider
dilute suspensions of small particles. Thus, 
effects of the particles on the fluid and particle-particle interactions are neglected (one
way coupling). 

The density of the particles was set to $\rho_p=1800 \rho$ where $\rho$ is the density of the fluid.
In each simulation the particles were divided in 24 fractions with different diameter $d$. The maximum diameter
equals the Kolmogorov length scale. Consequently, the particles can be treated as point particles.

The particle relaxation time
\beq
\tau_p=\frac{\rho_p}{\rho}\frac{d^2}{18 \nu}
\eeq
ranges from $\tau_p=0.1\tau_K$ to $\tau_p=100 \tau_K$.
Corresponding Stokes numbers $St=\frac{\tau_p}{\tau_K}$ based
on the Kolmogorov time scale $\tau_K$ range from $St=0.1$ to $St=100$ .

Based on the works of
\cite{citeulike:1387698} and \cite{Kubik04}, we assumed that 
in the given configurations
the acceleration of a particle $\frac{d {\bf v}}{dt}$ is given by Stokes drag only,
\beq
\frac{d {\bf v}}{dt} = -\frac{c_D Re_p}{24 \tau_p} ({\bf v} - {\bf u}_{f@p}).
\label{eqMaxey}
\eeq
Here, ${\bf v}(t)$ denotes the particle velocity and ${\bf u}_{f@p}$ the fluid velocity at the particle position.
The particle Reynolds number $Re_p$ is based on particle diameter and particle slip
velocity $\|{\bf u}_{f@p} - {\bf v}\|$ which leads to a nonlinear term for the Stokes drag. The drag coefficient 
$c_D$ was computed in dependence
of $Re_p$ according to the scheme proposed by \cite{Clift78}.

The fluid velocity ${\bf u}_{f@p}$ must be evaluated at the particle position ${\bf x}_p(t)$, i.e. ${\bf u}_{f@p} = {\bf
u}({\bf x}_p(t), 
t)$. Hence, these values must be interpolated.  
In the present work, a standard fourth order interpolation scheme was implemented, following the recommendations of
\cite{Yeung88} and \cite{Balachandar89}.

In the following,
the notation `$@p$' is adopted for arbitrary functions $f({\vf x},t)$, i.e.,
\beq
f_{@p}(t)=f\left({\vf{x}}_p\left(t\right),t\right).
\label{eqAtP}
\eeq
For example ${\vf{u}}_f$ refers to the space- and time-dependent solution of the Navier--Stokes equations whereas ${\vf
u}_{f@p}$ refers to the time-dependent fluid velocity seen by the particle. Correspondingly,
${\cal G} {\vf{u}}_f$ refers to the space- and time-dependent solution of the filtered Navier--Stokes equations whereas $\left( {\cal G} {\vf
u}_f\right)_{@p}$ refers to the time-dependent filtered fluid velocity seen by the particle.

Equation \eqref{eqMaxey} is a stiff differential equation
for small Stokes numbers. The numerical scheme for
integrating equation \eqref{eqMaxey} must be capable to handle this. Therefore, equation
\eqref{eqMaxey} was solved 
by a Rosenbrock-Wanner method \cite[see][]{Hairer90}. This method is a
fourth order method with adaptive time stepping. The stiff term in equation \eqref{eqMaxey} is
linearised in each time step and 
discretised by an implicit Runge-Kutta scheme.

\begin{figure}[H]
\includegraphics[bb=5 87 590 706, angle=-90,width=8.2cm]{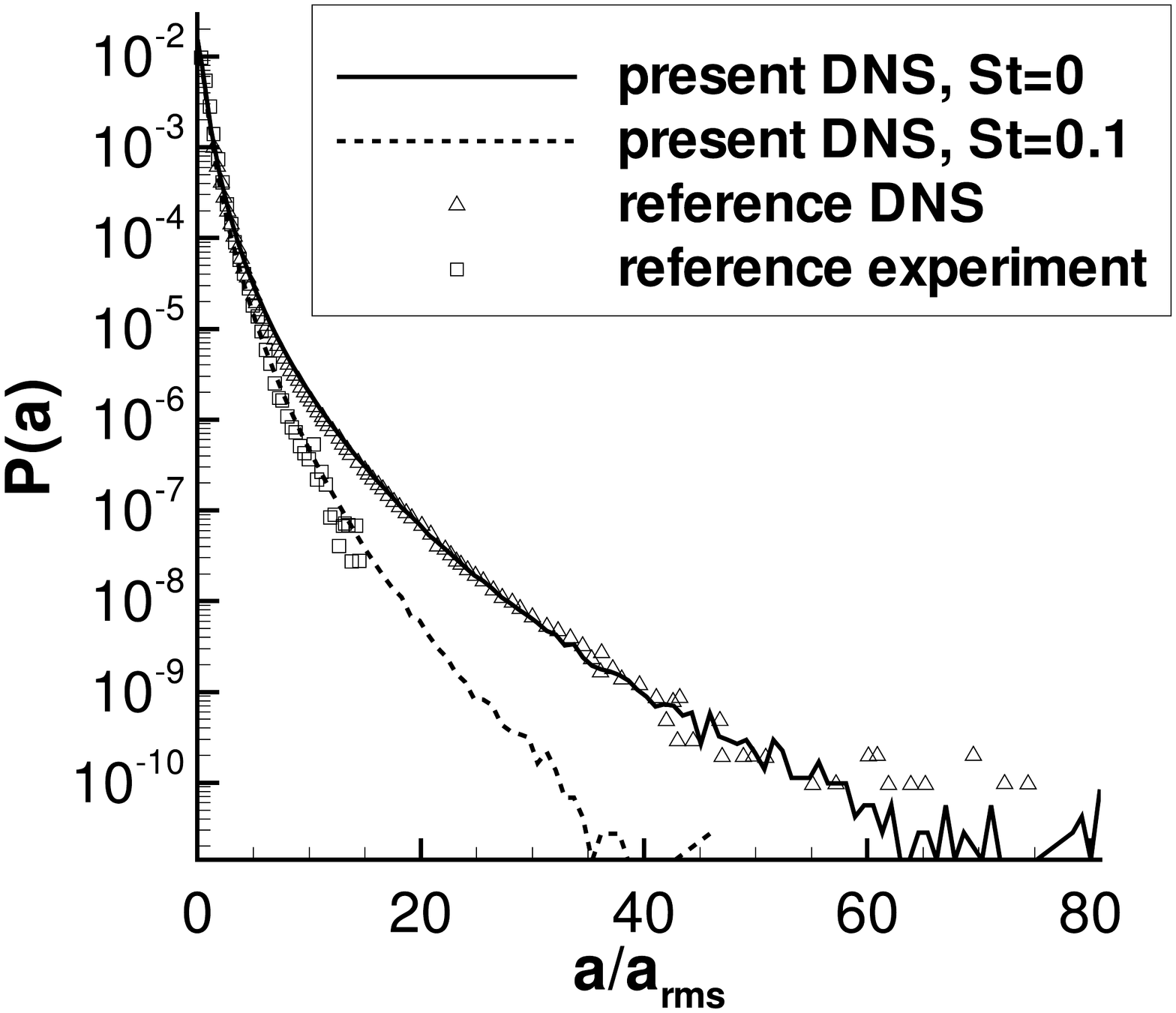}
\caption{\label{figAccelPDF}Probability density function $P({\bf a})$ of particle acceleration ${\bf a}$ for validation of the code. 
X-axis is normalised with respect to the (Stokes number dependent) rms value of ${\bf a}$.
Triangles: reference DNS of $St=0$ particles conducted by \cite{citeulike:5250032}. 
Squares: reference experiment of $St=0.09 \pm 0.03$ particles conducted by \cite{citeulike:5781002} (renormalised). 
Continuous and dashed lines: present DNS at
$St=0$ and $St=0.1$, respectively.}
\end{figure}

The code was validated via probability density functions (PDFs) for the particle acceleration. To this end, a DNS of forced isotropic turbulence
at $Re_\lambda=265$ on $1030^3$ grid points was conducted. This data was then compared to data from a DNS
conducted by \cite{citeulike:5250032} and an experiment conducted by \cite{citeulike:5781002}.
Biferale et al. conducted a DNS at $Re_\lambda=280$ and traced inertia free particles (i.e. $St=0$).
Ayyalasomayajula {\it et al.'s} experiment was 
at $Re_\lambda=250$ with particle Stokes numbers $St=0.09 \pm 0.03$.
Correspondingly, in the present simulation two particle fractions were traced, one at $St=0$ and another at $St=0.1$. 
Each fraction consists of 960000 particles.
Figure \ref{figAccelPDF} shows that the results from the present simulations
agree very well with the referenced data.

For model assessment,
24 fractions of particles were traced with 80000 particles per fraction. The particles were
initialised at random positions (homogeneous distribution)
inside the computational box and traced until a statistical steady state was obtained. Then, 
1000 time records were taken within a time span of $T = 250 \lambda/u_{rms}$ for computing statistics. 
The temporal resolution of the statistics equals approximately the Kolmogorov time scale.
With this temporal resolution, the
Lagrangian correlation functions could be resolved for all Stokes numbers. 
The time span was large enough to guarantee that
averaging in time cancels out oscillations caused by the forcing scheme.

In terms of particle time scales, $T$ is large enough to guarantee reliable statistics. From $\epsilon=15 \nu u_{rms}^2
/ \lambda^2$ it follows that $T/\tau_p = 250 \sqrt{15}/St \approx 968.2 / St$. In all simulations,
$St \le 100$, thus $T/\tau_p \ge 9.68$. Hence, statistics were sampled over at least 9.68 times the particle relaxation time.

\section{Analysed particle-LES models}

\label{secModels}

In the present section, the three particle-LES models under consideration are presented. They are 
the approximate deconvolution method which was proposed for particle-laden flow independently by  \cite{Kuerten06} and \cite{citeulike:1476706}
and two stochastic models, proposed by  \cite{citeulike:3191448} and \cite{Simonin93}. In the following, the models are stated and 
the numerical implementation used in this work is explained.

\subsection{Approximate Deconvolution Method (ADM)}

\paragraph{Model statement}
ADM is well established for incompressible single phase flows \cite*[see][]{citeulike:3821985,citeulike:3821986,citeulike:3821987}. 
\cite{citeulike:6129185,Kuerten06}, \cite{citeulike:1476706} and \cite{citeulike:3818015} 
analysed the capabilities of ADM for particle-laden flow.
With ADM, the fluid velocity seen by the particle ${\vf{u}}_{f@p}^{\mathit{ADM}}$ is computed from
\beq
{\vf{u}}_{f@p}^{\mathit{ADM}} = \left({\vf{u}}_{f}^{\mathit{ADM}}
\right)_{@p} = \sum\limits_{n=0}^N \left(\left({\cal I}-{\cal G}\right)^n {\cal G} {\vf{u}}_f
\right)_{@p} =  \left({\cal H}^{ADM} {\cal G} {\vf{u}}_f\right)_{@p} .
\label{eqADM}
\eeq
Here, ${\cal I}$ stands for identity. $N$ is the number of deconvolution steps. ${\cal H}^{ADM}$ is called defiltering
 operator because it is supposed to approximate the inverse of ${\cal G}$.

Equation \eqref{eqADM} is solved once per time step and the particle velocity is computed from
\beq
\sdf{{\vf{u}}_p^{\mathit{ADM}}}{t} = \frac{c_D Re_p}{24 \tau_p}\left({\vf{u}}_{f@p}^{\mathit{ADM}} - {\vf{u}}_p^{\mathit{ADM}}\right).
\eeq

The operator ${\cal H}={\cal I}-{\cal G}$ can be interpreted as extractor of subgrid scales.
With this operator, ${\vf{u}}_f^{\mathit{ADM}}$ can be written as
\beq
{\vf{u}}_f^{\mathit{ADM}} = \sum\limits_{n=0}^N {\cal H}^n {\cal G} {\vf{u}}_f = \sum\limits_{n=0}^N {\cal H}^n \left({\cal I} - {\cal
H}\right) {\vf{u}}_f = 
\left({\cal I} - {\cal H}^{N+1}\right) {\vf{u}}_f.
\label{eqIdADM}
\eeq
For $N\rightarrow \infty$ the transfer function of ${\cal H}^{N+1}$ equals zero for the resolvable scales \linebreak
($\|{\vf k}\|<\kappa_c$) and one for the unresolvable scales ($\|{\vf k}\|>\kappa_c$).
 This shows that for large $N$, 
the effect of ADM can be interpreted as improving the LES filter towards a sharp spectral filter. 

\paragraph{Implementation of the model in this work}
In the present work, the ADM defiltering operator ${\cal H}^{ADM}$
was computed in three different ways. First, it was computed as proposed by \cite{Kuerten06}. Second, it was computed 
making use of the DNS spectrum and third, a model spectrum was used.

\begin{figure}[H]
\includegraphics[bb= 100 0 700 590, clip, width=8.2cm]{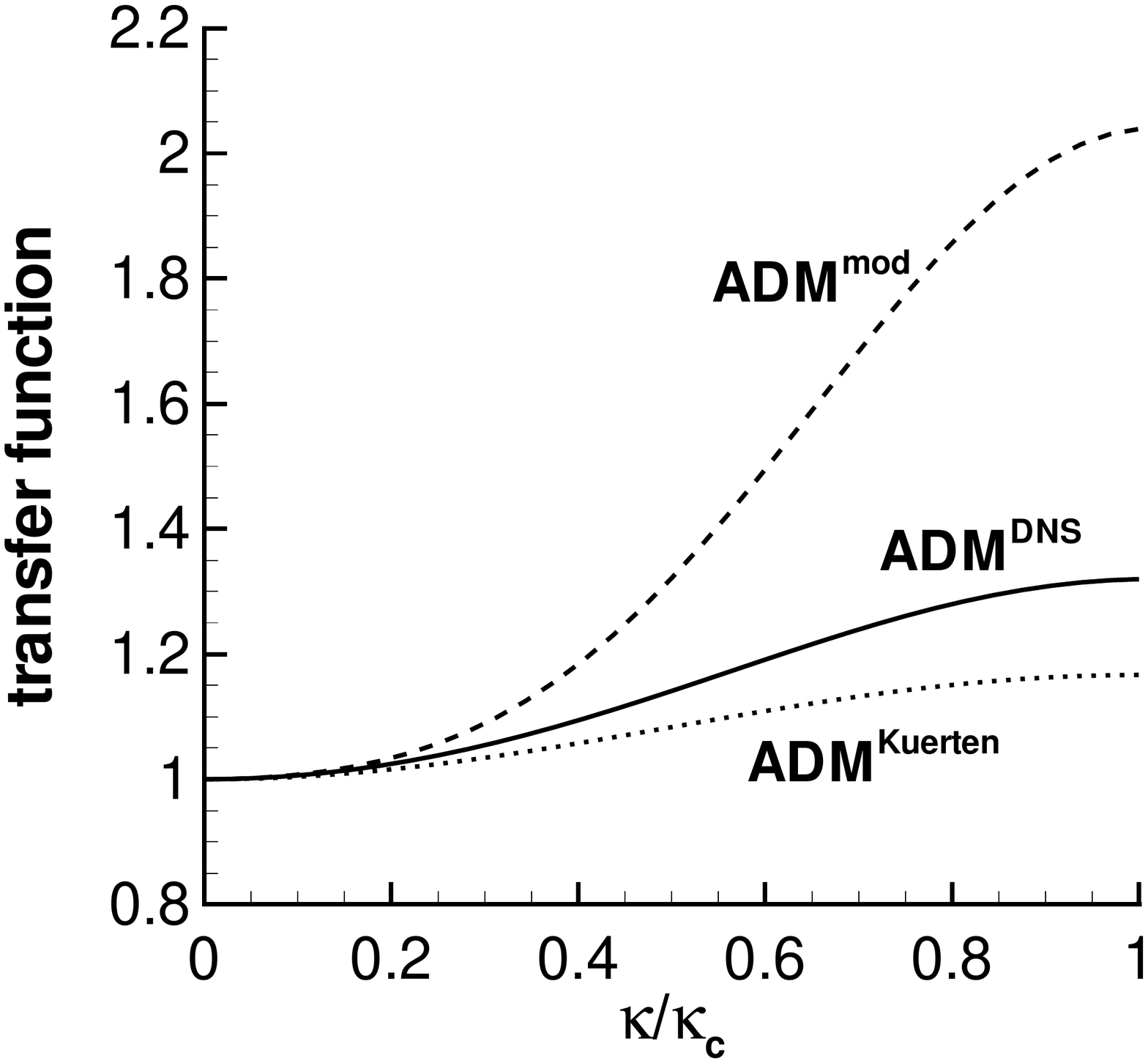}
\caption{\label{figTFRe50Modspec}Transfer functions of the defiltering operators for the three implemented ADM approaches.}
\end{figure}

If a dynamic Smagorinsky
model is used as fluid-LES model, then
\cite{Kuerten06} proposes to compute ${\cal H}^{ADM}$ as approximate inverse of the corresponding test filer. In his work and in the present work, this is a box filter.
\cite{Kuerten06} approximates its inverse by a second-order Taylor expansion in 
the filter width. 
The transfer function of this filter is shown in figure \ref{figTFRe50Modspec}.
In the following this approach is referred to as ADM\super{Kuerten}.

However, it is not clear whether an inverted box filter gives highest accuracy. Therefore ADM was tested by two more approaches. In both approaches,
the ADM filter is constructed such that the product of filter transfer function and LES spectrum is as close as possible to a target spectrum under the constraint
that the filter stencil covers up to $5^3$ LES cells. The target spectrum is either the DNS spectrum or the model spectrum proposed by \cite{citeulike:1666191}. 
The results from the
corresponding defiltering operators are referred to as ADM\super{DNS} and ADM\super{mod}, respectively. 
The corresponding transfer functions are also shown in figure \ref{figTFRe50Modspec}.
Evidently ADM\super{mod} leads to a very much stronger amplification around $\kappa_c$ than ADM\super{DNS}. 
This was to be expected because around $\kappa_c$ the model spectrum is higher than the DNS spectrum, cf. figure \ref{figSpectr}.

ADM does not take explicitly into account that the model itself affects the particle path.
More precisely, the model inherently assumes that the resolved
spectrum seen by the particle is not modified by the model itself. 
In order to differentiate between this model assumption
and other approximation errors of the model, the a priori analysis 
was conducted such that the model does not affect the particle path.

More precisely, in the a priori analysis for ADM, 
for each particle two different values for the particle velocity were computed simultaneously. 
One value, referred to as DNS particle velocity, is the velocity obtained from the DNS flow field. 
The second value, referred to as modelled particle velocity, is the velocity obtained from a filtered DNS field and ADM. 
The particles were tracked with the DNS particle velocity 
and statistical samples were taken from the modelled velocity. 
This approach basically tests whether ADM is capable to do what it is supposed to do, neglecting the effect of ADM on the particle path.

\subsection{Langevin-based models proposed by Shotorban and Mashayek and Simonin et al.}
\label{secModelSho}

ADM cannot reconstruct scales smaller than the LES grid. In order to circumvent this,
\cite{citeulike:3191448} and \cite{Simonin93} propose stochastic models based on a Langevin
equation for the fluid velocity seen by a particle. 
Such models were originally developed for inertia free particles by
\cite{citeulike:3821994}, \cite{Heinz03} and \cite{Gicquel02}, referred to as generalised Langevin models. 

\paragraph{Statement of the model proposed by \cite{citeulike:3191448}}
Shotorban and Mashayek adopted generalised Langevin models for inert particles. They propose to compute 
the fluid velocity seen by the particles ${\vf{u}}_{f@p}^{\mathit{Sho}}$ from the stochastic differential equation (Langevin equation)
\beq
\intd u_{f@p,i}^{\mathit{Sho}} = \left( {\cal G} \left(\df{u_{f,i}}{t} + u_{f,j} \df{u_{f,i}}{x_j}\right) \right)_{@p} \intd t 
- \frac{u_{f@p,i}^{\mathit{Sho}}-\left( {\cal G} u_{f,i} \right)_{@p} }{T_L} \intd t + \sqrt{C_0 \epsilon} \; \intd W_i
\label{eqShotorban}
\eeq
and the particle velocity from
\beq
\intd {\vf{u}}_p^{\mathit{Sho}} = \frac{c_D Re_p}{24 \tau_p}\left({\vf{u}}_{f@p}^{\mathit{Sho}} - {\vf{u}}_p^{\mathit{Sho}}\right) \intd t.
\eeq
The reader is reminded that `$@p$' denotes `at the particle position', cf. equation \eqref{eqAtP}.
The first term on the right hand side of equation \eqref{eqShotorban} is the filtered 
material derivative of the fluid velocity and can be computed from the right hand side of the
filtered Navier--Stokes equation. 
The second term is a drift term for the random variable ${\vf{u}}_{f@p}^{\mathit{Sho}}$, leading to a
relaxation of ${\vf{u}}_{f@p}^{\mathit{Sho}}$ against $\left( {\cal G}{\vf{u}}_f \right)_{@p}$. The
last term is a diffusion term for ${\vf{u}}_{f@p}^{\mathit{Sho}}$. ${\vf{W}}$ denotes a Wiener process and $\epsilon$ is the (modelled)
dispersion of subgrid scale kinetic energy. The model parameters $T_L$ and $C_0$ are specified below.

\paragraph{Statement of the model proposed by \cite{Simonin93}}
\cite{Simonin93} also propose to model the fluid velocity seen by the particles by a stochastic process. 
\cite{citeulike:1666260} presented in detail how to deduct Simonin et al.'s model 
for particle-laden flow starting from the Navier--Stokes
equations. 
This results in a different Langevin equation than the equation proposed by Shotorban and Mashayek. 

In contrast to Shotorban and Mashayek,
Simonin et al. propose to transport
the resolved scales by particle velocity (and not by fluid velocity).
The model can be formulated via a
Langevin equation for the unresolved scales
\beq
\intd u_{f@p,i}^{{\mathit{Sim}}^\prime} = \left( -u_{f@p,j}^{{\mathit{Sim}}^\prime} \left(\df{{\cal G}
u_{f,i}}{x_j} \right)_{@p}
+ \left(\df{\tau_{i,j}}{x_j}\right)_{@p}
+ \Gamma_{ij} u_{f@p,j}^{{\mathit{Sim}}^\prime} \right) \intd t + \sqrt{C_0 \epsilon} \; \intd W_i.
\label{eqSimonin}
\eeq
$\tau_{ij} = {\cal G}\left(u_i u_j\right) - {\cal G}u_i {\cal G}u_j$ is the SGS stress tensor.
The model constant $C_0$ is equivalent
to $C_0$ of Shotorban and Mashayek's model.
For isotropic turbulence, \cite{citeulike:1666260} recommend
\beq
\Gamma_{ij}=-\frac{\left(\frac{1}{2} + \frac{3}{4} C_0\right)\epsilon}{k_{sgs}} \; \delta_{ij} = -\frac{1}{T_L} \; \delta_{ij} ,
\label{eqModelGamma}
\eeq
$\delta_{ij}$ denoting the Kronecker delta function. This form of $\Gamma$ was adopted in the present work.

The fluid velocity seen by the particles is then computed 
from ${\vf{u}}_{f@p}^{\mathit{Sim}}=\left( {\cal G} {\vf{u}}_f \right)_{@p}
 + {\vf{u}}_{f@p}^{\mathit{Sim}^\prime}$
and the particle velocity is computed from
\beq
\intd {\vf{u}}_p^{\mathit{Sim}} = \frac{c_D Re_p}{24 \tau_p}\left({\vf{u}}_{f@p}^{\mathit{Sim}} - {\vf{u}}_p^{\mathit{Sim}}\right) \intd t.
\eeq

\paragraph{Implementation of the model in this work}
For model closure, two parameters need to be specified, namely the time scale $T_L$ and the Kolmogorov constant $C_0$. Based on the recommendation of the model's authors,
in the present work $T_L$ was set to
\beq
T_L=\frac{k_{sgs}}{\left(\frac{1}{2} + \frac{3}{4} C_0\right)\epsilon}, \qquad \epsilon=C_\epsilon\frac{k^{3/2}}{\Delta}.
\label{eqModelTL}
\eeq
$k_{sgs}$ denotes the subgrid kinetic energy and was computed from the DNS data. 
$\epsilon$ denotes the SGS rate of dispersion.
The model constants $C_\epsilon$ and 
$C_0$ were set to $C_\epsilon=1$ and $C_0=2.1$, following \cite{citeulike:3817867} and \cite{Gicquel02}. 

For the model of \cite{Simonin93},
the SGS stress tensor $\tau$ was computed in accordance with the fluid-LES model, i.e., 
using an eddy viscosity hypothesis. 

As mentioned above, for the a priori analysis of ADM, the particles were traced along the path computed from DNS.
For the stochastic models, this would be in contradiction to the model assumptions because the 
model takes explicitly into account that the particle
path depends on the modelled small scale fluctuations \cite[see][]{citeulike:1476706, citeulike:1666260}. Therefore here the particle paths were computed from the
modelled fluid velocity.

The stochastic differential equations \eqref{eqShotorban} and \eqref{eqSimonin} 
were solved by an Euler-Maruyama scheme 
\cite[see e.g.][]{citeulike:1220909}. The stiff terms $-u_{f@p,i}^{\mathit{Sho}}/T_L$ and $\Gamma_{ij} u_{f@p,j}^{{\mathit{Sim}}^\prime}$ 
 were discretised implicitly. 
 \cite{citeulike:1476706} and \cite{citeulike:1666260} used an explicit Euler-Maruyama scheme. These authors focussed on small Stokes numbers.
In the present simulations no significant differences between explicit and implicit discretisation was found at
small Stokes numbers.  At high Stokes numbers, the explicit scheme was found to produce significantly worse results. In particular, the 
kinetic energy seen by the particles explodes at high Stokes numbers when using an explicit scheme.
It should be noted that the terms under consideration are linear and therefore implicit
schemes do not produce any computational overhead.
In the following, `Sho' denotes results from the model proposed by \cite{citeulike:1476706} and
`Sim' denotes results for the model proposed by \cite{Simonin93}.

\section{A priori and a posteriori analysis of particle-LES models} 
\label{secNumAssess}

The present section contains results from a priori and a posteriori analysis for assessment of the three
models presented in section \ref{secModels}. The analysis comprises the kinetic energy seen by the particles, particle kinetic energy and rate of dispersion.

\subsection{Assessment of ADM}
\label{secNumAssessADM}

\begin{figure}[!b]
\begin{minipage}[b]{0.48\linewidth}
\includegraphics[bb=70 5 700 540, clip,width=\linewidth]{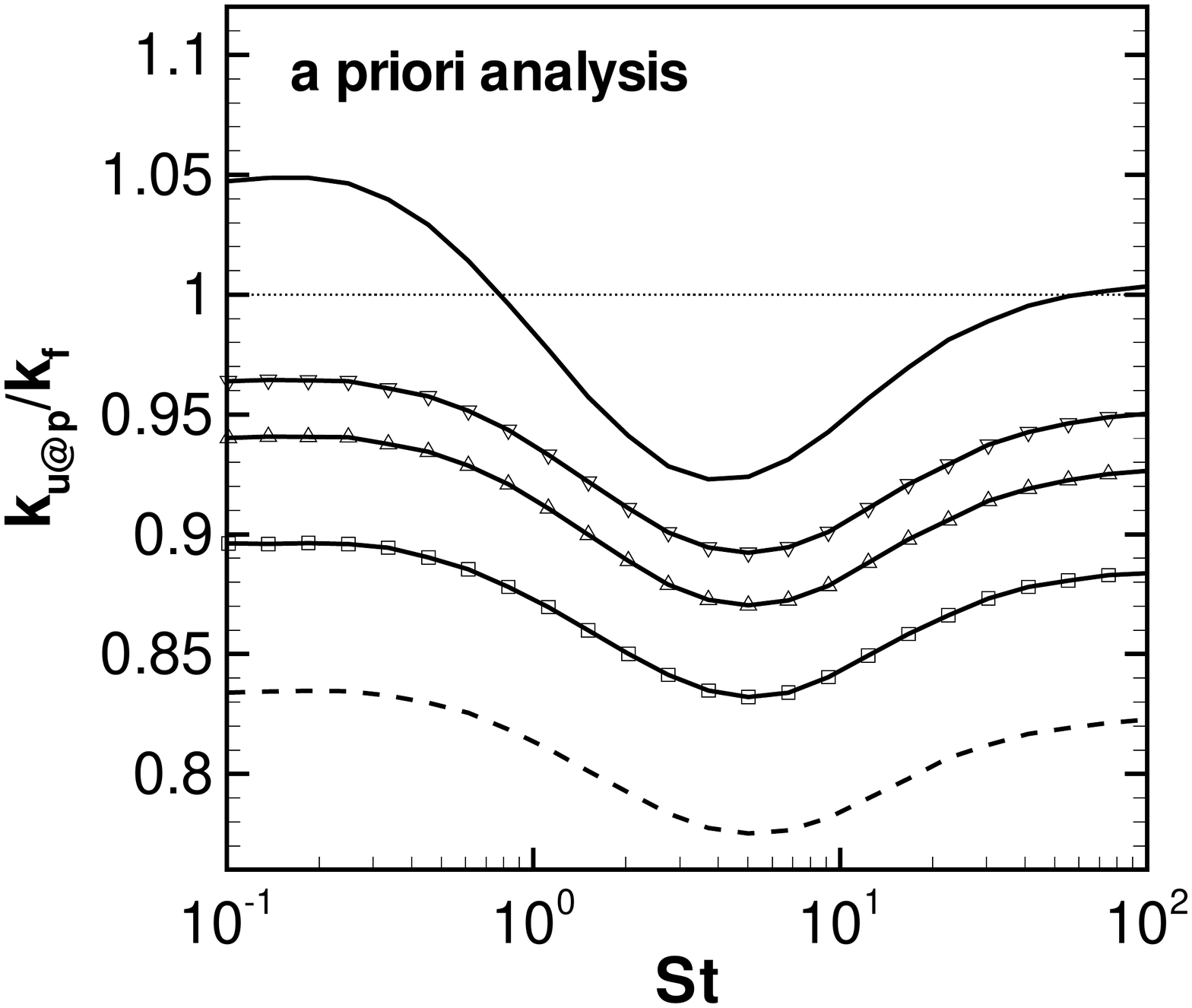}
\vspace{0pt}
\end{minipage}
\hfill
\begin{minipage}[b]{0.48\linewidth}
\includegraphics[bb=70 5 700 540, clip,width=\linewidth]{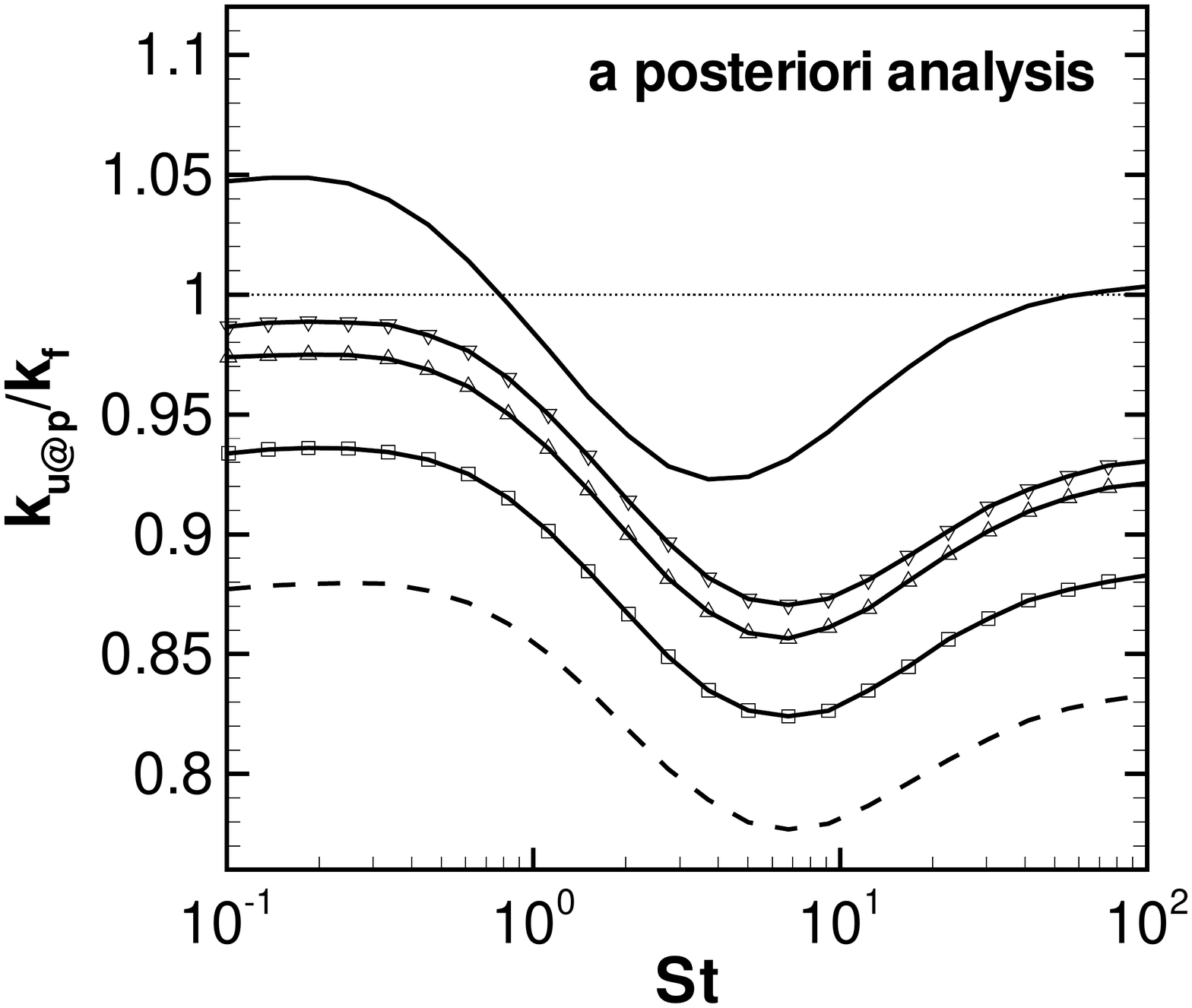}
\vspace{0pt}
\end{minipage}
\centering
\includegraphics[bb=107 685 185 725, clip,width=0.056\linewidth]{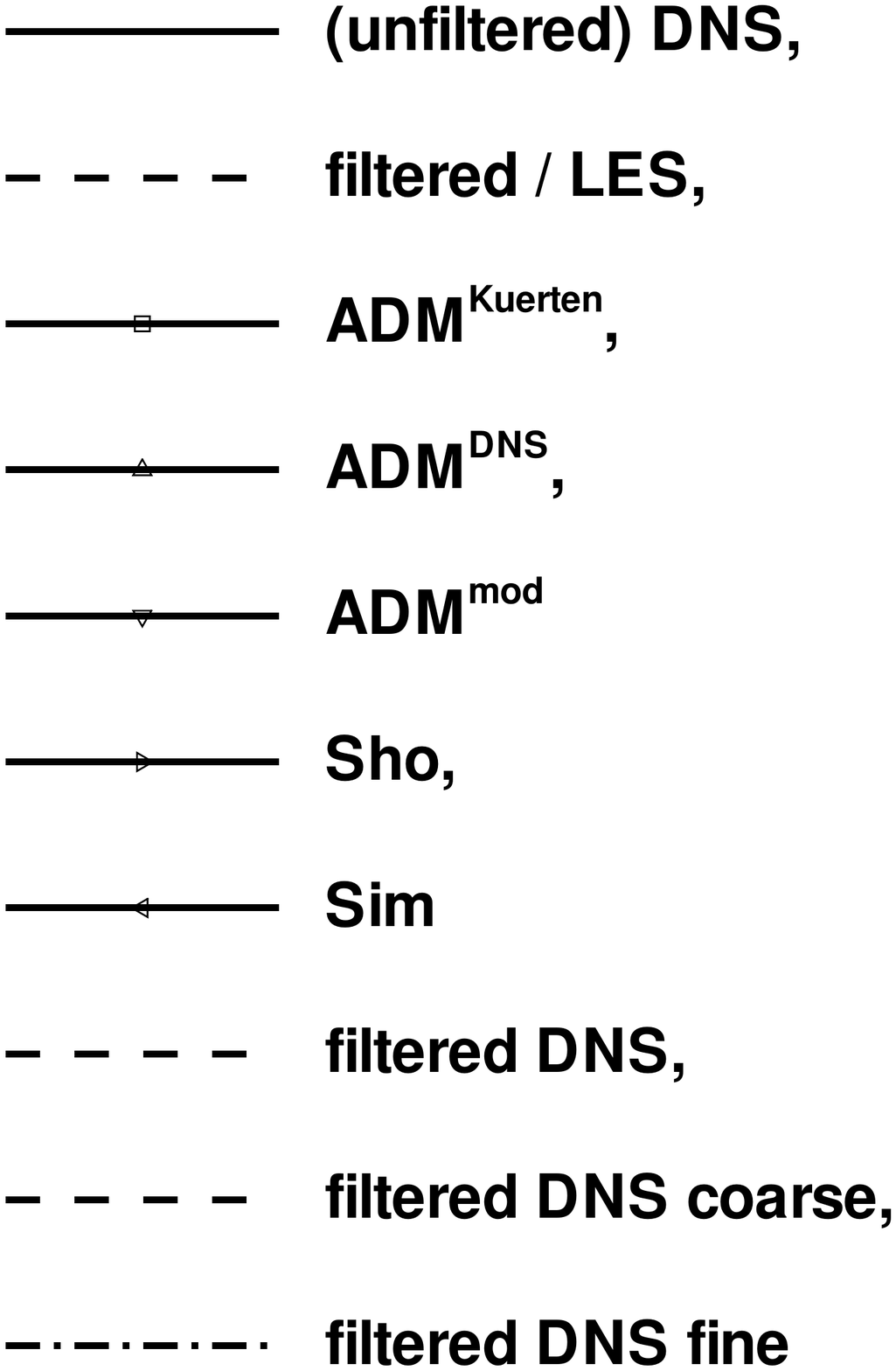}~
\includegraphics[bb=225 685 469 725, clip,width=0.177\linewidth]{legend.pdf}~
\includegraphics[bb=107 618 185 658, clip,width=0.056\linewidth]{legend.pdf}~
\includegraphics[bb=225 618 415 658, clip,width=0.137\linewidth]{legend.pdf}~
\includegraphics[bb=107 551 185 591, clip,width=0.056\linewidth]{legend.pdf}~
\includegraphics[bb=225 551 380 591, clip,width=0.11\linewidth]{legend.pdf}~
\includegraphics[bb=107 484 185 524, clip,width=0.056\linewidth]{legend.pdf}~
\includegraphics[bb=225 484 355 524, clip,width=0.094\linewidth]{legend.pdf}~
\includegraphics[bb=107 416 185 456, clip,width=0.056\linewidth]{legend.pdf}~
\includegraphics[bb=225 416 355 456, clip,width=0.094\linewidth]{legend.pdf}
\caption{\label{figADMKineF}
A priori (left) and a posteriori (right) analysis of ADM, kinetic energy seen by particles.}
\end{figure}

\begin{figure}[!b]
\begin{minipage}[b]{0.48\linewidth}
\includegraphics[bb=70 5 700 540, clip,width=\linewidth]{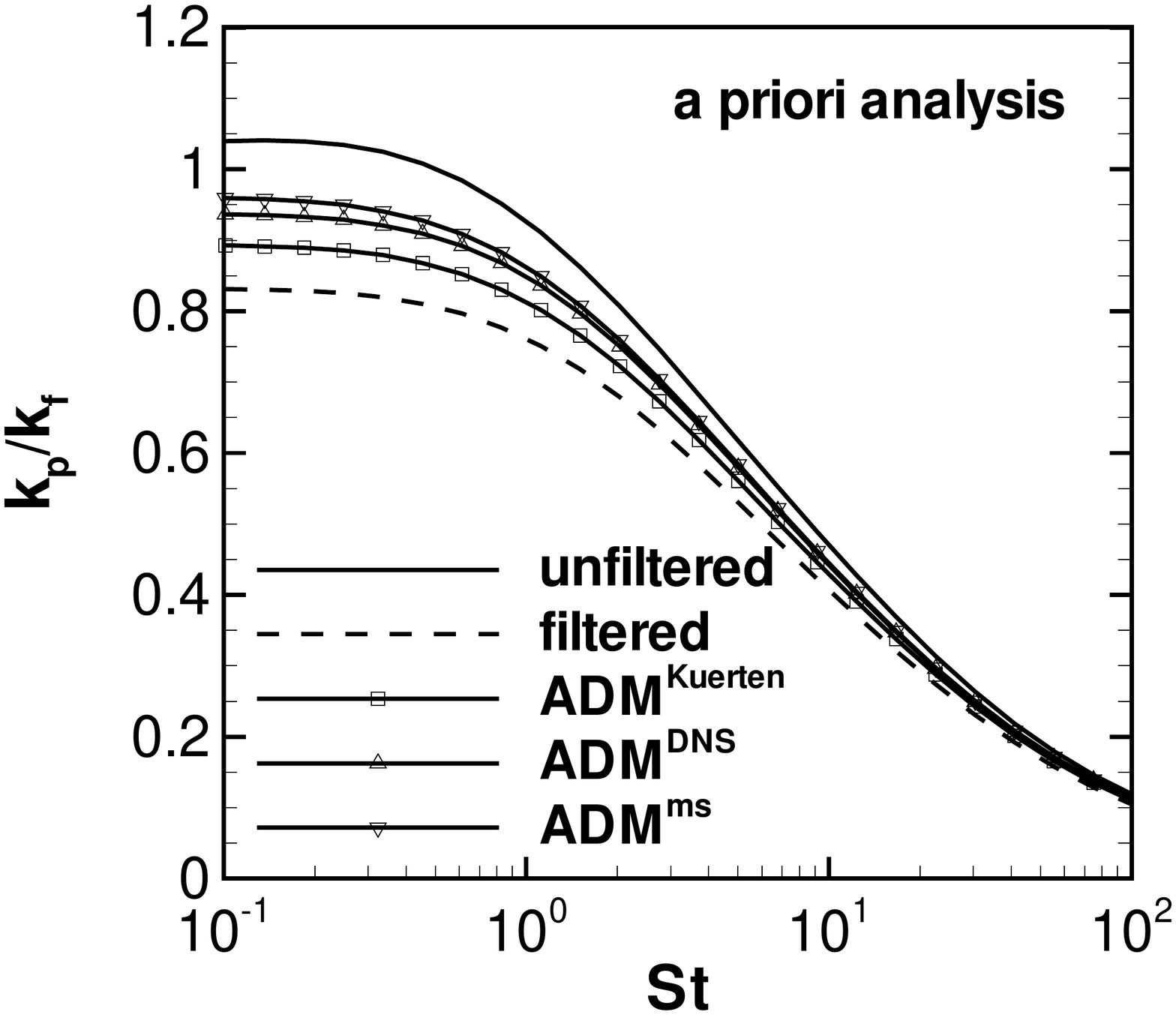}
\vspace{0pt}
\end{minipage}
\hfill
\begin{minipage}[b]{0.48\linewidth}
\includegraphics[bb=70 5 700 540, clip,width=\linewidth]{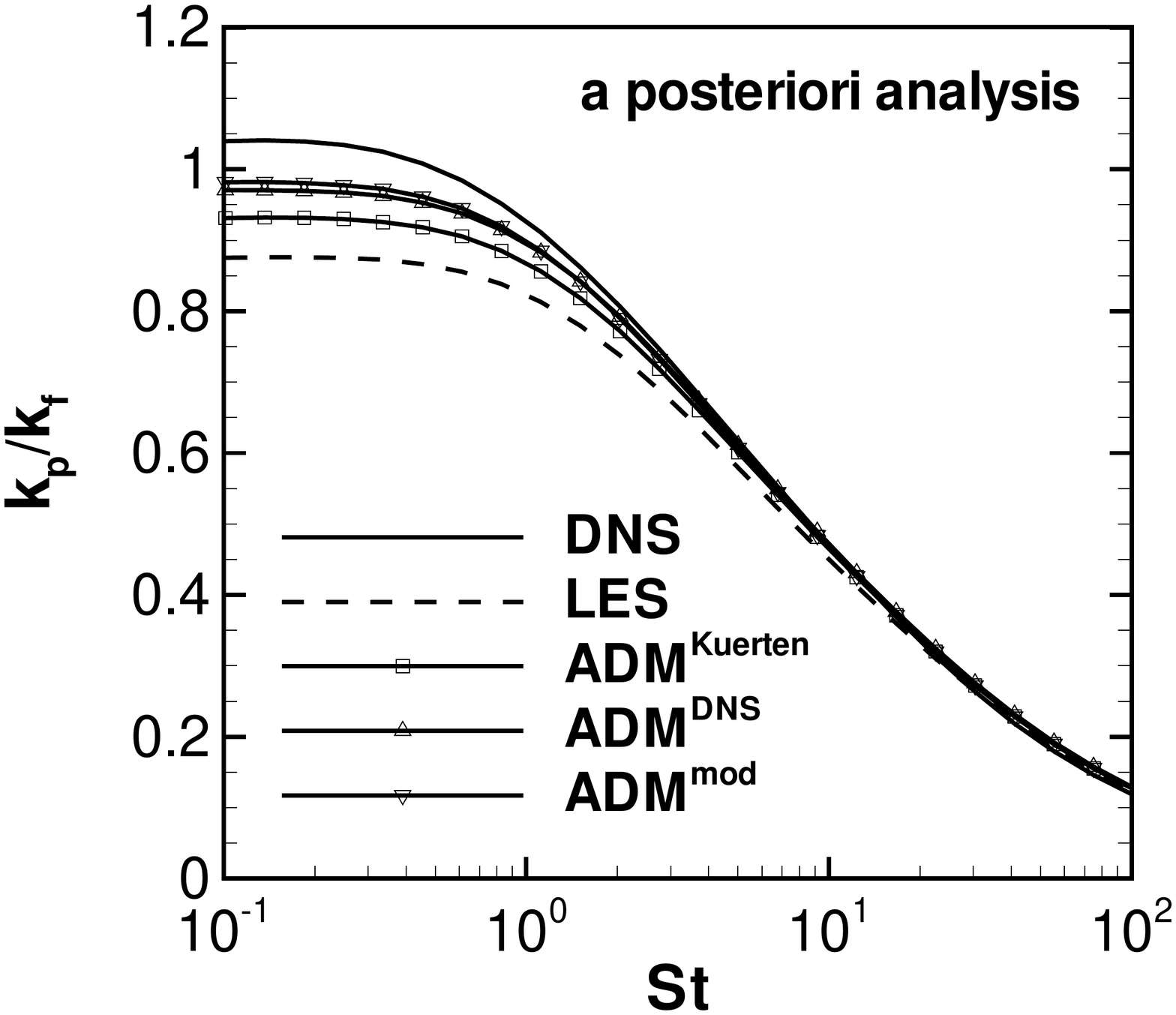}
\vspace{0pt}
\end{minipage}
\caption{\label{figADMKine}
A priori (left) and a posteriori (right) analysis of ADM, particle kinetic energy.}
\end{figure}

\begin{figure}[!b]
\begin{minipage}[b]{0.48\linewidth}
\includegraphics[bb=70 5 700 540, clip,width=\linewidth]{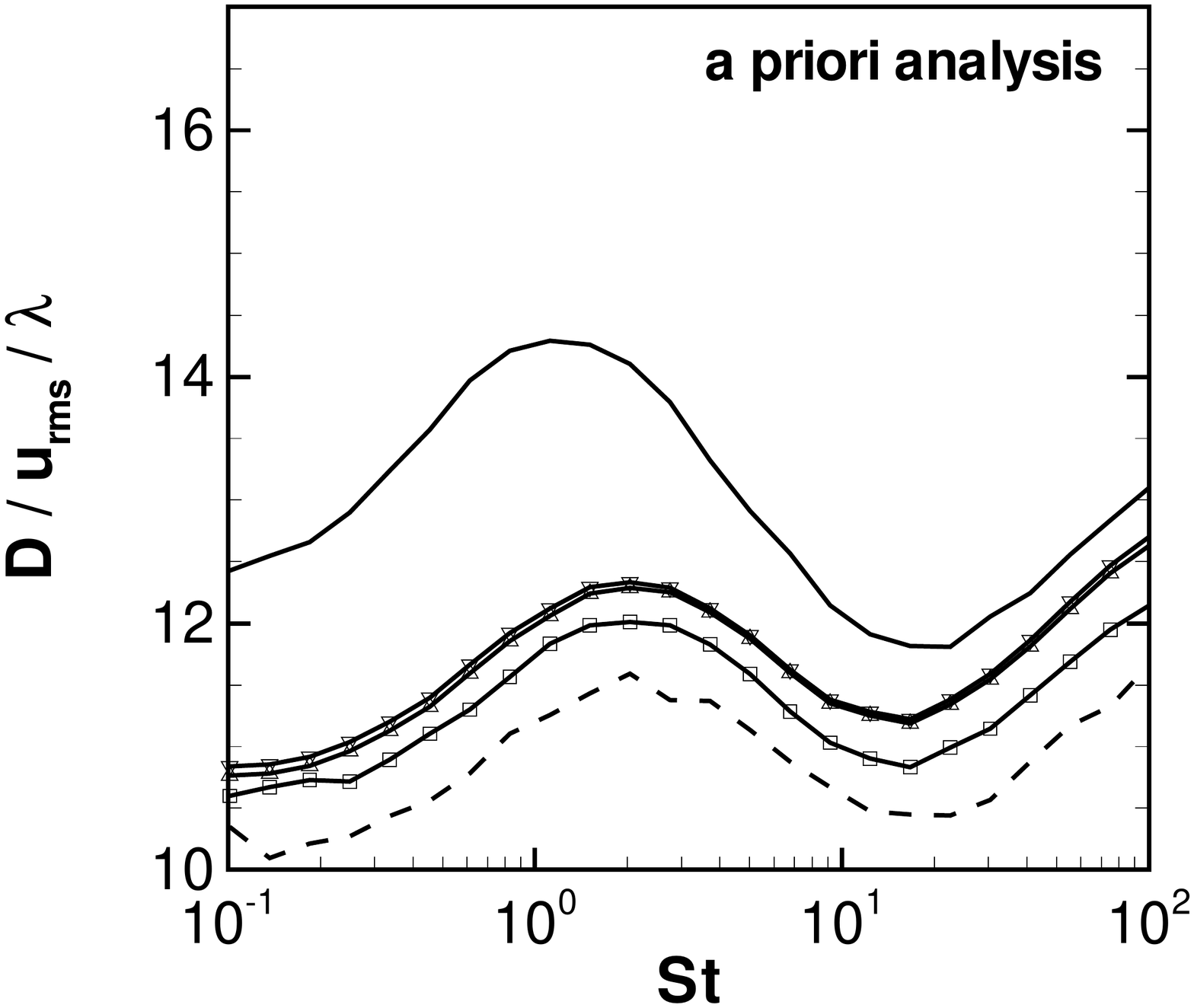}
\vspace{0pt}
\end{minipage}
\hfill
\begin{minipage}[b]{0.48\linewidth}
\includegraphics[bb=70 5 700 540, clip,width=\linewidth]{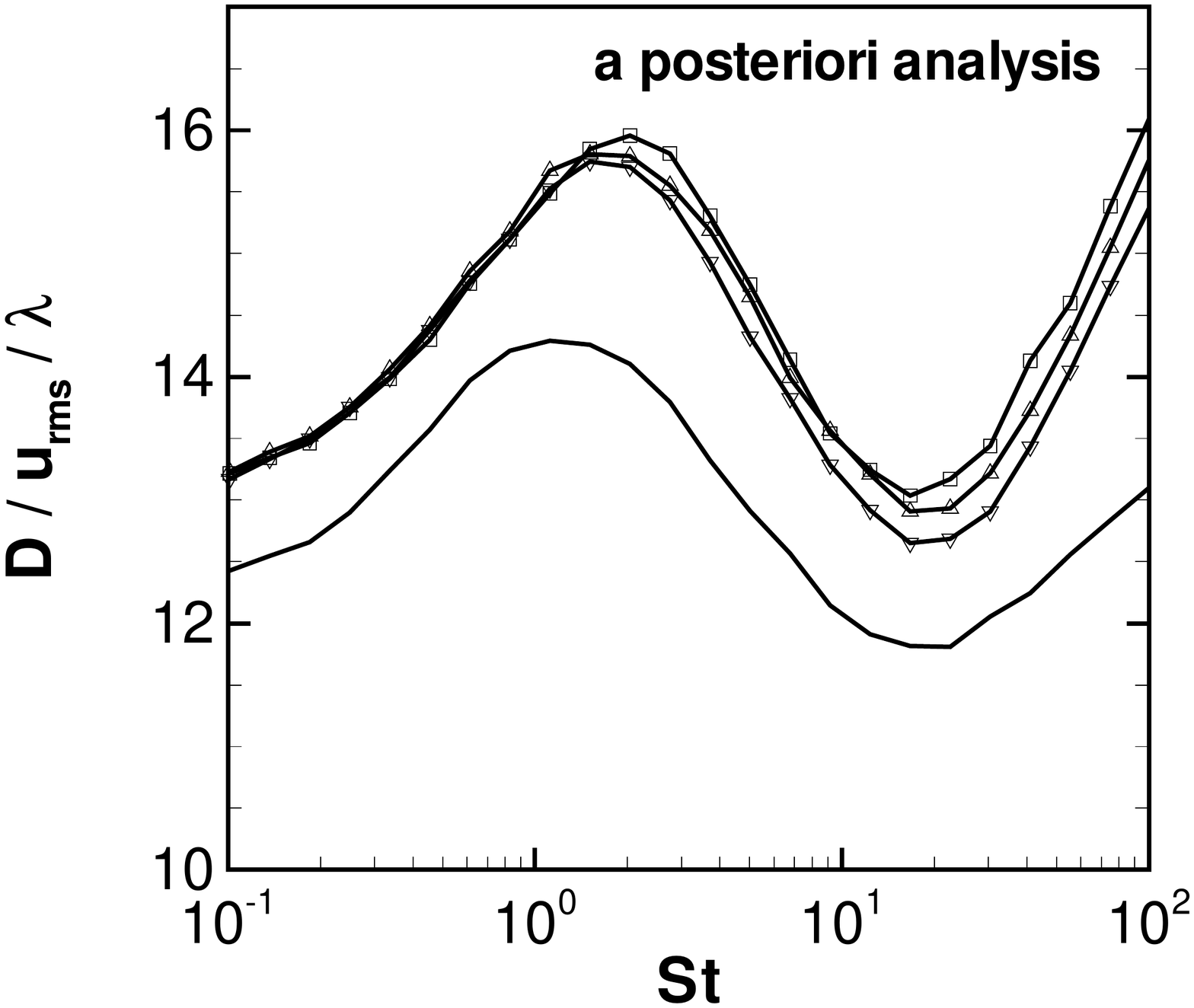}
\vspace{0pt}
\end{minipage}
\centering
\includegraphics[bb=107 685 185 725, clip,width=0.056\linewidth]{legend.pdf}~
\includegraphics[bb=225 685 469 725, clip,width=0.177\linewidth]{legend.pdf}~
\includegraphics[bb=107 215 185 255, clip,width=0.056\linewidth]{legend.pdf}~
\includegraphics[bb=225 215 410 255, clip,width=0.134\linewidth]{legend.pdf}~
\includegraphics[bb=107 551 185 591, clip,width=0.056\linewidth]{legend.pdf}~
\includegraphics[bb=225 551 380 591, clip,width=0.11\linewidth]{legend.pdf}~
\includegraphics[bb=107 484 185 524, clip,width=0.056\linewidth]{legend.pdf}~
\includegraphics[bb=225 484 355 524, clip,width=0.094\linewidth]{legend.pdf}~
\includegraphics[bb=107 416 185 456, clip,width=0.056\linewidth]{legend.pdf}~
\includegraphics[bb=225 416 355 456, clip,width=0.094\linewidth]{legend.pdf}
\caption{\label{figADMPrioDispers}
A priori (left) and a posteriori (right) analysis of ADM, rate of dispersion.
Result from LES without particle-LES model (not shown for reasons of clarity) is almost identical to the results from LES with ADM.}
\end{figure}

Figures \ref{figADMKineF} 
to \ref{figADMPrioDispers} show the kinetic energy seen by the particles, particle kinetic energy and the rate of dispersion
for the three ADM implementations under consideration.
In addition, results from filtered DNS and LES without particle-LES model are shown. 
In order to obtain comparable results, 
the presented data from filtered DNS without model corresponds to the data which ADM receives, i.e.,
in particular in the filtered DNS the particles were traced along unfiltered paths.

The kinetic energy seen by the particles (figure \ref{figADMKineF}) shows
a Stokes number dependence due to particle clustering. The Stokes number dependence can be observed in all simulations although
the results from filtered DNS and LES are shifted towards higher Stokes numbers in comparison to the unfiltered DNS.
This shift is not corrected by ADM.

As expected, the kinetic energy seen by the particles is lower in LES or filtered DNS than in unfiltered DNS.
ADM leads to a clear improvement of the kinetic energy seen by the particles although a clear gap between results from ADM and DNS persists. 
This gap was to be expected because ADM does not reconstruct the smallest scales.

Among the three ADM approaches,
Kuerten's model shows least improvement. This is not surprising because Kuerten's approach
corresponds to a single defiltering step, $N=1$. The other two approaches must perform better here because the wider
stencils allow for larger values of 
$N$, leading to higher kinetic energy. On the other hand it must be mentioned that ADM\super{Kuerten} is computationally less
expensive than ADM\super{DNS} or ADM\super{mod} and somewhat more generalistic in the sense that ADM\super{Kuerten} 
only assumes a specific filter whereas ADM\super{DNS} and ADM\super{mod} are based on the spectra of this specific flow.

On first sight, the comparison of ADM\super{DNS} against ADM\super{mod} is surprising.
ADM\super{mod}  shows closer resemblance to the unfiltered result than ADM\super{DNS} although ADM\super{DNS} is based on the unfiltered field. 
Actually this is an effect of two errors cancelling out each other. 
Around the cutoff wavenumber
the model spectrum is higher than the DNS spectrum but interpolation of the fluid velocity on the particle position leads to strong damping
around the cutoff wavenumber. In other words, for ADM\super{mod}
the damping properties of the
interpolation scheme 
brings the spectrum seen by the particles
closer to the spectrum of the DNS flow field.

Figure \ref{figADMKine} shows $k_u$, the kinetic energy of the particles themselves. 
As expected from the observations on $k_{u@p}$, $k_u$ is underestimated by ADM. With $St\rightarrow \infty$ the
error vanishes.

Most interesting is the rate of dispersion, shown in figure \ref{figADMPrioDispers}. 
It was computed from the product of particle kinetic energy and integral time scale.
The results from a priori and a posteriori analysis differ \emph{qualitatively}. 
In the a priori analysis, ADM leads to an underprediction 
of the rate of dispersion for all Stokes numbers. 
The reader is reminded that in the a priori analysis of ADM, the particles were traced along DNS particle paths.

The a posteriori analysis shows that actually in LES the rate of dispersion is overestimated by ADM as a result of too high integral time scales. 
In the a posteriori analysis of ADM, the particles were traced with the modelled particle velocity.
Thus,
the qualitative difference between a priori analysis and a
posteriori analysis could be twofold; either due to the difference in particle paths or due to a 
defect in the fluid-LES model. Additional simulations showed that the principal reason
is that the fluid-LES model leads to too high life times for the large eddies and thus to an overprediction of particle dispersion.
Consequently, it depends on the fluid-LES model whether the rate of dispersion is under- or overestimated by LES.

However, a priori and a posteriori analysis show that even for high Stokes numbers where small scale effects should be negligible \cite[see][]{Yamamoto01},
the rate of dispersion is not predicted correctly by ADM.

\subsection{Assessment of the Langevin-based models}
\label{secNumAssessLang}

\begin{figure}[!b]
\begin{minipage}[b]{0.48\linewidth}
\includegraphics[bb=70 5 700 540, clip,width=\linewidth]{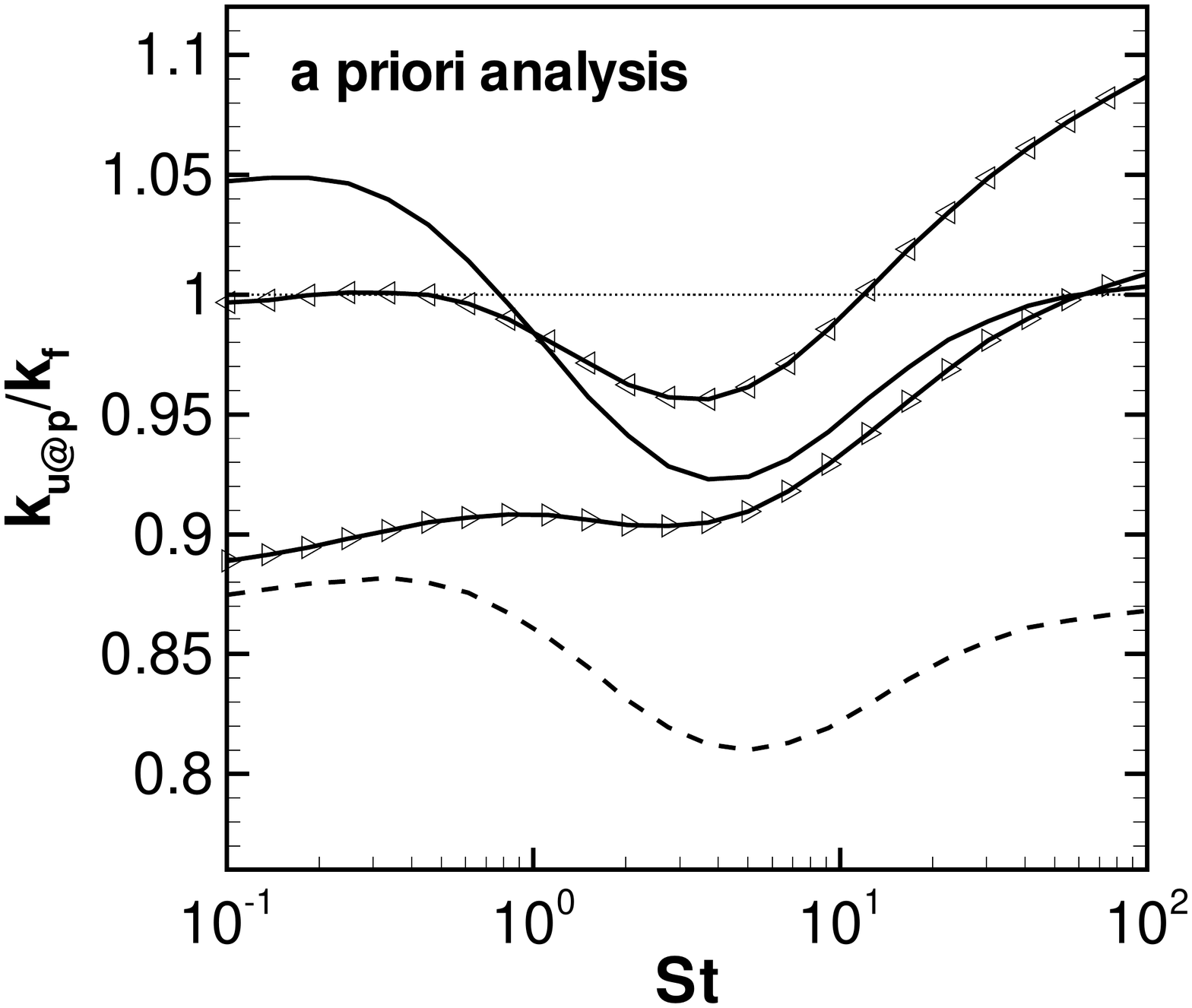}
\vspace{0pt}
\end{minipage}
\hfill
\begin{minipage}[b]{0.48\linewidth}
\includegraphics[bb=70 5 700 540, clip,width=\linewidth]{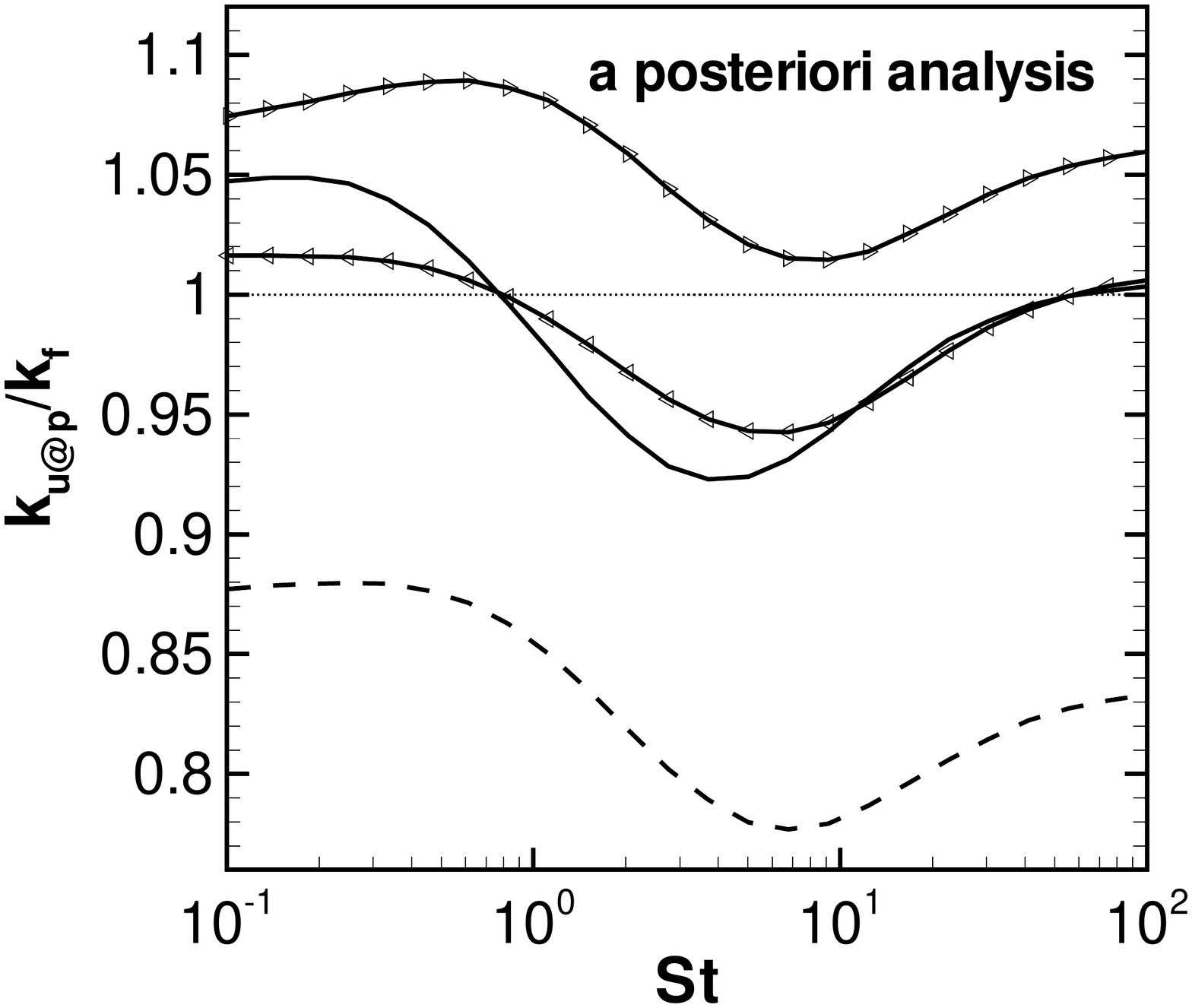}
\vspace{0pt}
\end{minipage}
\centering
\includegraphics[bb=107 685 185 725, clip,width=0.056\linewidth]{legend.pdf}~
\includegraphics[bb=225 685 469 725, clip,width=0.177\linewidth]{legend.pdf}~
\includegraphics[bb=107 618 185 658, clip,width=0.056\linewidth]{legend.pdf}~
\includegraphics[bb=225 618 415 658, clip,width=0.137\linewidth]{legend.pdf}~
\includegraphics[bb=107 349 185 389, clip,width=0.056\linewidth]{legend.pdf}~
\includegraphics[bb=225 349 305 389, clip,width=0.058\linewidth]{legend.pdf}~
\includegraphics[bb=107 282 185 322, clip,width=0.056\linewidth]{legend.pdf}~
\includegraphics[bb=225 282 305 322, clip,width=0.058\linewidth]{legend.pdf}
\caption{\label{figShoSimKineF} 
A priori (left) and a posteriori (right) analysis of the Langevin-based models, kinetic energy seen by particles.
}
\vspace{-5ex}
\end{figure}

The testcase for the Langevin-based models is again isotropic turbulence at $Re_\lambda=52$.
Figure \ref{figShoSimKineF} shows the kinetic energy of the fluid seen by the particles. In particular the results from the a priori analysis are very disappointing. 
The model of \cite{citeulike:3191448} shows too little $k_{u@p}$ for $St=0.1$ although $k_{sgs}$ was computed from the DNS data.
We found that this is an effect of the convective term in the model, $\left( {\cal G} \left(\df{u_{f,i}}{t} + u_{f,j} \df{u_{f,i}}{x_j}\right) \right)_{@p}$.
If this term is neglected, then $k_{u@p}$ is correctly predicted by the model.

Concerning the a priori analysis, for low Stokes numbers 
the model of \cite{Simonin93} leads to a more accurate prediction of $k_{u@p}$
than the model of \cite{citeulike:3191448} but for high Stokes numbers one can observe the inverse.
In the a posteriori analysis, the model of \cite{Simonin93} gives a satisfactory prediction for $k_{u@p}$
whereas the model of \cite{citeulike:3191448} leads to an overestimation.

\begin{figure}[!b]
\begin{minipage}[b]{0.48\linewidth}
\includegraphics[bb=70 5 700 540, clip,width=\linewidth]{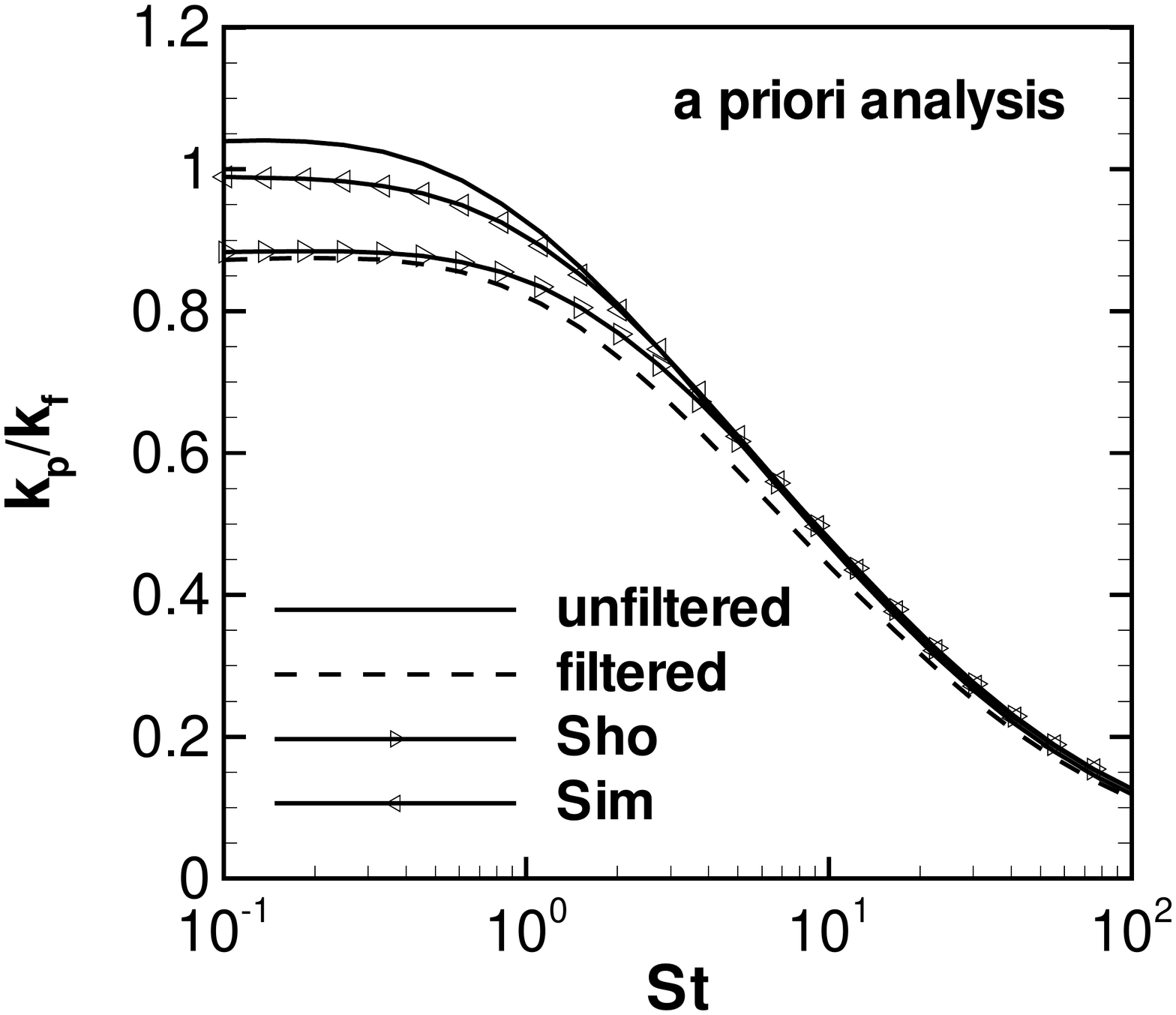}
\vspace{0pt}
\end{minipage}
\hfill
\begin{minipage}[b]{0.48\linewidth}
\includegraphics[bb=70 5 700 540, clip,width=\linewidth]{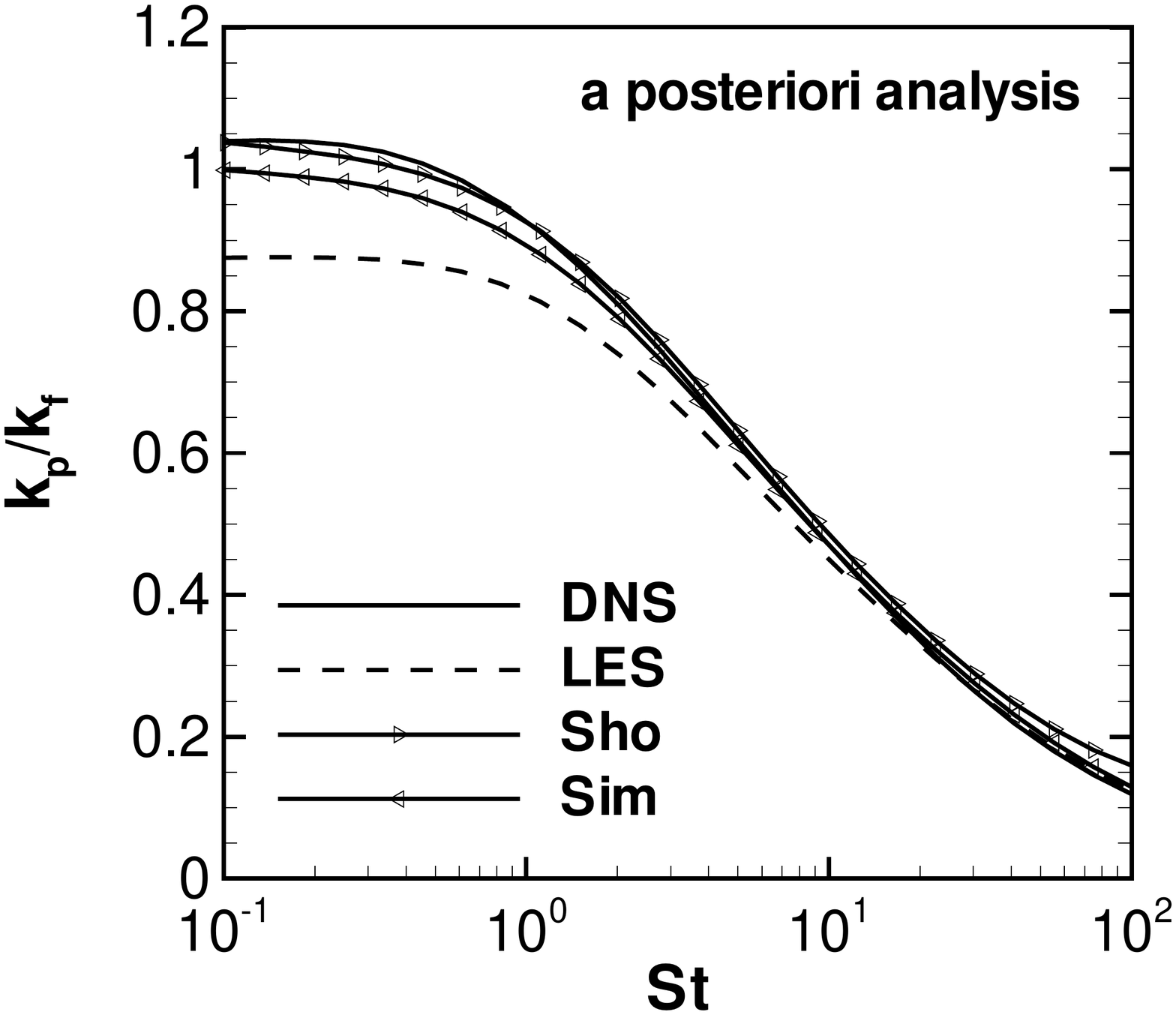}
\vspace{0pt}
\end{minipage}
\caption{\label{figShoSimKine}
A priori (left) and a posteriori (right) analysis of the Langevin-based models, particle kinetic energy.
}
\end{figure}

The error in the kinetic energy of the particles is simply a consequence of the error in the kinetic energy seen by the particles, cf. figure \ref{figShoSimKine}. 
In the a priori analysis of the model of \cite{Simonin93}, the excess in $k_{u@p}$ for high $St$ is not apparent in $k_u$.
On the other hand, in the a posteriori analysis of the model of \cite{citeulike:1476706}, the excess in $k_{u@p}$ for high $St$ is also visible in $k_u$.
This is probably due to the different time stepping. In LES, the time step size
is higher than in DNS. For the particles $St=100$, the ratio between particle relaxation time and time step size is about 1000 for DNS
and about 200 for LES. Thus, the $St=100$-particles can rather follow the modelled fluctuations in LES than in DNS.
Therefore, the excess in $k_{u@p}$ at high Stokes number is rather reflected in the a posteriori analysis than in the a priori analysis.

\begin{figure}[H]
\begin{minipage}[b]{0.48\linewidth}
\includegraphics[bb=70 5 700 540, clip,width=\linewidth]{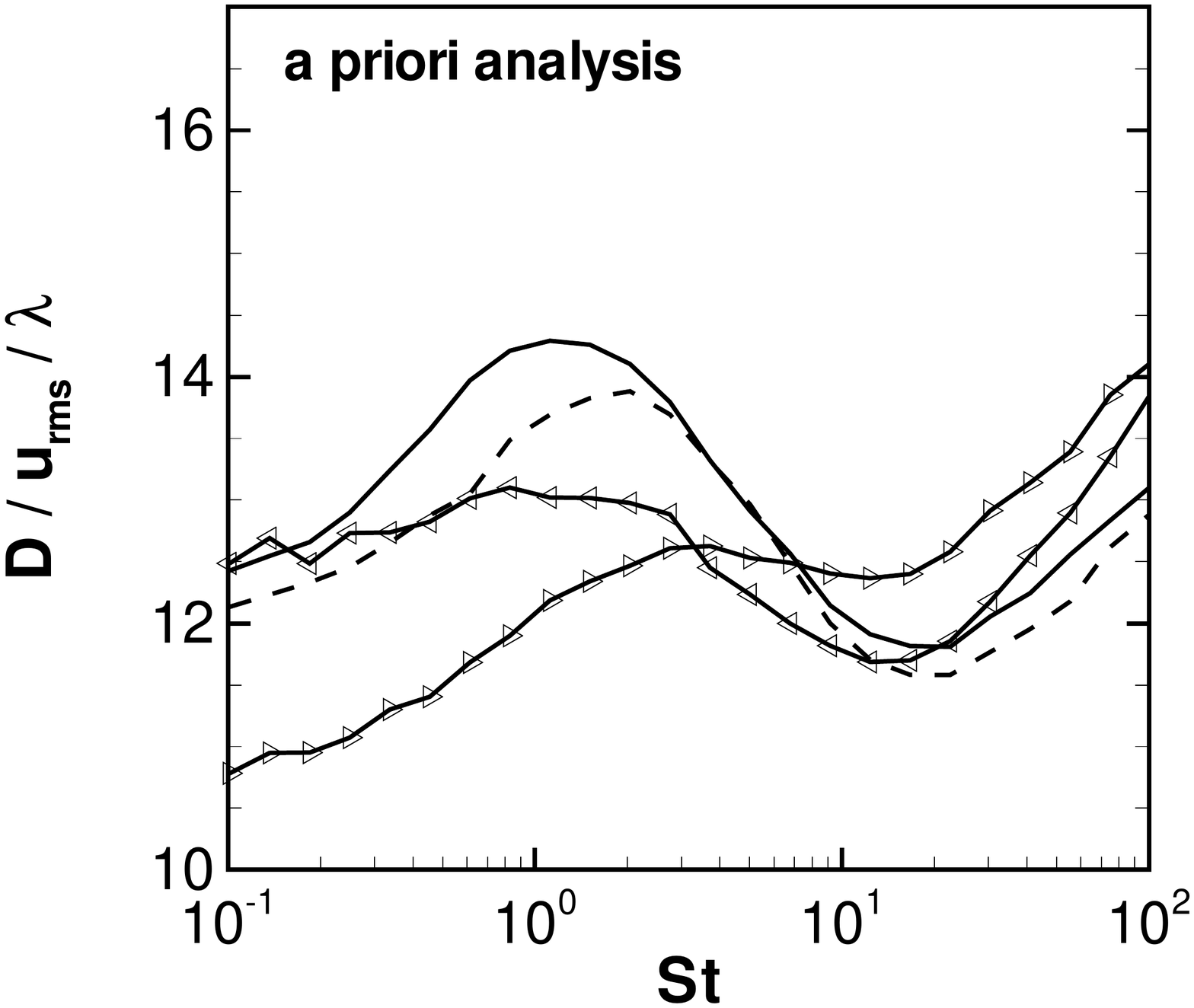}
\vspace{0pt}
\end{minipage}
\hfill
\begin{minipage}[b]{0.48\linewidth}
\includegraphics[bb=70 5 700 540, clip,width=\linewidth]{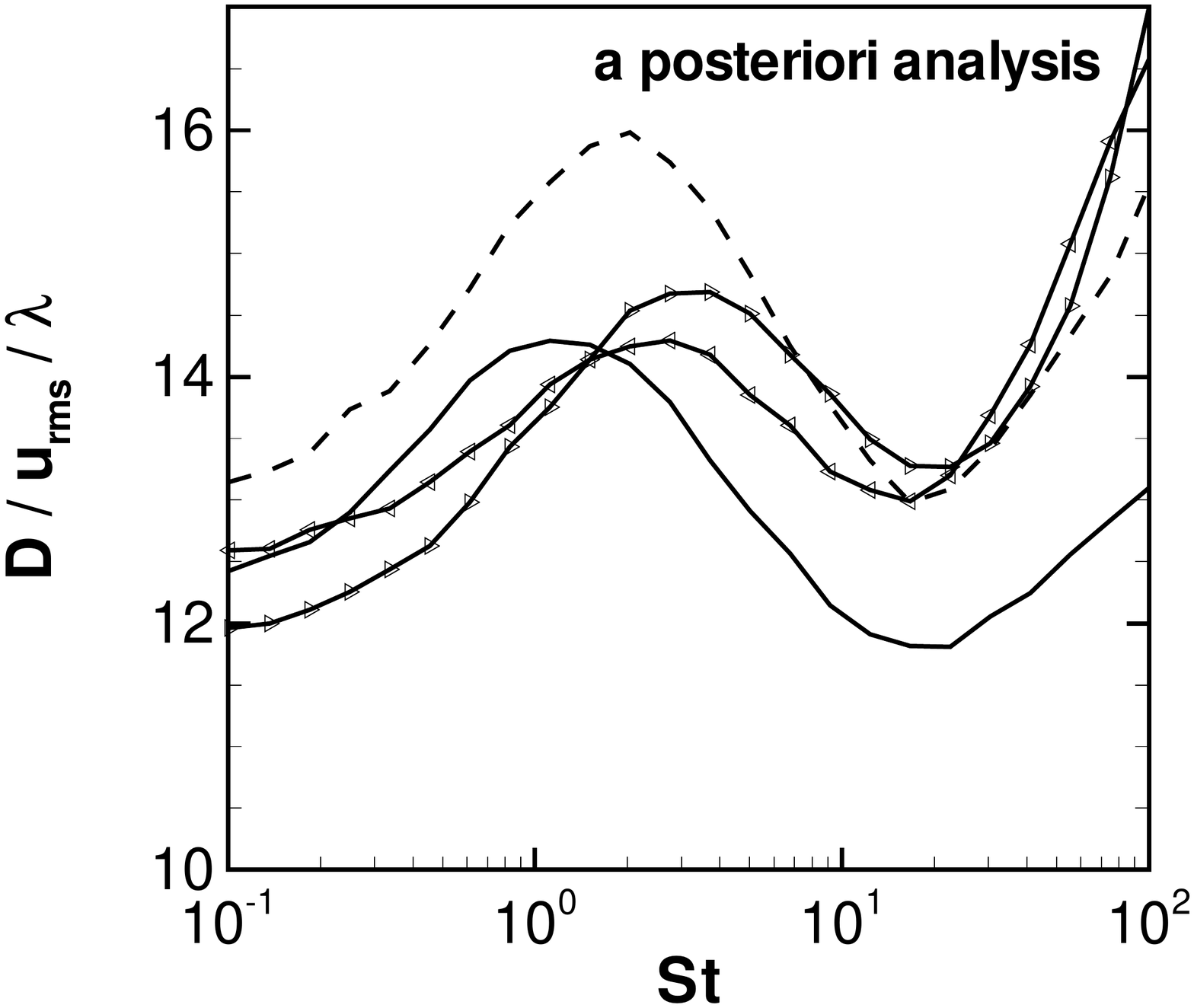}
\vspace{0pt}
\end{minipage}
\centering
\includegraphics[bb=107 685 185 725, clip,width=0.056\linewidth]{legend.pdf}~
\includegraphics[bb=225 685 469 725, clip,width=0.177\linewidth]{legend.pdf}~
\includegraphics[bb=107 618 185 658, clip,width=0.056\linewidth]{legend.pdf}~
\includegraphics[bb=225 618 415 658, clip,width=0.137\linewidth]{legend.pdf}~
\includegraphics[bb=107 349 185 389, clip,width=0.056\linewidth]{legend.pdf}~
\includegraphics[bb=225 349 305 389, clip,width=0.058\linewidth]{legend.pdf}~
\includegraphics[bb=107 282 185 322, clip,width=0.056\linewidth]{legend.pdf}~
\includegraphics[bb=225 282 305 322, clip,width=0.058\linewidth]{legend.pdf}
\caption{\label{figShoSimPrioDispers}
A priori (left) and a posteriori (right) analysis of the Langevin-based models, rate of dispersion.
}
\end{figure}

Concerning the rate of dispersion, figure \ref{figShoSimPrioDispers},
the result from the a priori analysis is very discouraging, the a posteriori results are somewhat better concerning accuracy of the models.
Nevertheless, these tests show that both models do not necessarily improve the result of the LES in comparison to an LES without particle-LES model or LES with ADM.
In particular at high Stokes numbers it might be recommendable to use no model instead of one of the stochastic models, in accordance with the numerical results of
\cite{citeulike:1476706}.

\section{Conclusions}

\label{secConclusions}

We have presented data from DNS, filtered DNS and LES of particle-laden isotropic turbulence 
with three different models for the effect of the unresolved scales on the particles (particle-LES models).
The models under consideration are 
approximate deconvolution (ADM) as proposed by  \cite{Kuerten06} and
two stochastic models, based on the works of \cite{citeulike:3191448}
and \cite{Simonin93}.
The present work allows for the first time a direct comparison of these models because all models
were assessed on the same testcase.

The models were assessed by a priori and a posteriori analysis. 
The analyses comprise the kinetic energy seen
by the particles, particle kinetic energy and rate of dispersion.

ADM was implemented in three different ways, one following \cite{Kuerten06} and two approaches via optimisation against spectra.
With all three approaches a defect in the rate of dispersion
could be observed. ADM was found to underestimate the rate of dispersion if applied
on the filtered DNS field. In LES however, ADM was found to overpredict the rate of dispersion. This discrepancy was explained as an effect
of approximation errors in the fluid-LES model.

The stochastic models showed very poor performance in the numerical simulations. In particular for high Stokes numbers, LES or filtered DNS with stochastic
models showed larger difference to DNS results than LES or filtered DNS without particle-LES model.

According to these results, the stochastic models are not recommendable because, in dependence of the configuration,
LES without particle-LES model can perform better, sometimes even tremendously
better, than LES with a stochastic particle-LES model. On the other hand, results from ADM are quite promising. At least ADM was found to lead to an improvement 
for all Stokes numbers.

On the other hand, ADM only enhances the resolved scales but does not actually model scales which cannot be represented on the grid.
At high Reynolds numbers, where the LES grid is very coarse due to computational limitations, ADM can be expected to perform worse
than at low Reynolds numbers. Therefore a new particle-LES model is needed for LES at high Reynolds numbers.

\bibliography{c-gobert}
\bibliographystyle{model2-names}

\end{document}